\documentclass[aps,prd,twocolumn,superscriptaddress]{revtex4}
\usepackage{epsfig,epsf}
\usepackage{amsmath}
\usepackage{amsthm}
\usepackage{amsfonts}
\usepackage{amssymb}
\usepackage{dsfont}
\usepackage{multirow}
\usepackage{appendix}
\usepackage{slashed}
\usepackage[active]{srcltx}
\usepackage{psfrag}

\begin{document}

\title{Decays of fully beauty scalar tetraquarks to $B_{q}\overline{B}_{q}$
and $B_{q}^{\ast}\overline{B}_{q}^{\ast}$ mesons}
\date{\today}
\author{S.~S.~Agaev}
\affiliation{Institute for Physical Problems, Baku State University, Az--1148 Baku,
Azerbaijan}
\author{K.~Azizi}
\thanks{Corresponding Author}
\affiliation{Department of Physics, University of Tehran, North Karegar Avenue, Tehran
14395-547, Iran}
\affiliation{Department of Physics, Do\v{g}u\c{s} University, Dudullu-\"{U}mraniye, 34775
Istanbul, T\"{u}rkiye}
\author{B.~Barsbay}
\affiliation{Division of Optometry, School of Medical Services and Techniques, Do\v{g}u%
\c{s} University, 34775 Istanbul, T\"{u}rkiye}
\author{H.~Sundu}
\affiliation{Department of Physics Engineering, Istanbul Medeniyet University, 34700
Istanbul, T\"{u}rkiye}

\begin{abstract}
Decays of the fully beauty four-quark structures $X_{\mathrm{4b}}$ and $T_{%
\mathrm{4b}}$ to $B$ meson pairs are investigated in the framework of the QCD
three-point sum rule method. We model the scalar exotic mesons $X_{\mathrm{4b%
}}$ and $T_{\mathrm{4b}}$ as diquark-antidiquark systems composed of the
axial-vector and pseudoscalar diquarks, respectively. The masses $m=(18540
\pm 50)~\mathrm{MeV}$ and $\widetilde{m}=(18858 \pm 50)~\mathrm{MeV}$ of
these compounds calculated in our previous articles, fix possible decay
channels of these particles. In the present work, we consider their decays
to $B_{q}\overline{B}_{q}$ and $B_{q}^{\ast }\overline{B}_{q}^{\ast }
(q=u,d,s,c)$ mesons. In the case of $X_{\mathrm{4b}}$ the mass of which is
below the $2\eta_{b}$ threshold, these channels determine essential part of
its full width $\Gamma_{\mathrm{4b}}$. The tetraquark $T_{\mathrm{4b}}$ can
decay to the pair $\eta_{b}\eta_{b}$, therefore partial widths of processes
with $B (B^{\ast})$ mesons in the final state permit us to refine our
estimate for the full width of this particle. The predictions $\Gamma_{%
\mathrm{4b}}=(9.6\pm 1.1)~\mathrm{MeV}$ and $\widetilde{\Gamma }_{\mathrm{4b}%
}^{\mathrm{Full}}=(144 \pm 29)~\mathrm{MeV}$ obtained in this article can be
used in future experimental investigations of four $b$-quark mesons.
\end{abstract}

\maketitle

\section{Introduction}

\label{sec:Intro}
%%%%%%%%%%%%%%%%%%%%%%%%%%%%%%%%%%%%%%%%%%%%%%%%%%%%%%%%%%%
Interest in  four-quark exotic mesons containing heavy $c$ and $b$ quarks
appeared in the first years of the parton model and QCD \cite%
{Iwasaki:1975pv,Chao:1980dv,Ader:1981db,Heller:1985cb,Lipkin:1986dw,Zouzou:1986qh}%
. Close attention to these hypothetical particles was inspired by many
reasons. First of all, fundamental laws of  QCD allow the existence of
multiquark hadrons, therefore such states became objects for intensive
theoretical studies. The second reason was a possibility to find multiquark
particles stable against strong decays, hence with a long  mean lifetime.
Investigations showed that tetraquarks, i.e., four-quark mesons built of a
heavy $bb$ diquark and light antidiquark may have desired features. Such
candidates to strong-interaction stable particles were analyzed in various
publications by means of different models and methods (see Refs.\ \cite%
{Karliner:2017qjm,Eichten:2017ffp,Agaev:2018khe,Agaev:2020mqq} and
references therein).

Fully heavy tetraquarks were also considered in numerous papers aimed to
reveal their properties. Recent data of the LHCb-ATLAS-CMS Collaborations
provided new experimental information \cite%
{LHCb:2020bwg,Bouhova-Thacker:2022vnt,CMS:2023owd}, which is important for
physics of heavy exotic mesons. These experiments discovered four $X$
resonances in the invariant mass distributions of the di-$J/\psi $ and $%
J/\psi \psi ^{\prime }$ mesons. The $X$ particles have masses in the range $%
6.2-7.3~\mathrm{GeV}$ and presumably are fully charmed states, though
alternative explanations were suggested as well.

In our articles \cite%
{Agaev:2023wua,Agaev:2023ruu,Agaev:2023gaq,Agaev:2023rpj}, we studied the $X$
structures as fully charmed scalar particles using both the
diquark-antidiquark and hadronic molecule models. We calculated their masses
and full width by employing the QCD two- and three-point sum rule (SR)
methods, and compared obtained results with the LHCb-ATLAS-CMS data. In
accordance with our predictions, the resonance $X(6600)$ is the tetraquark
composed of axial-vector diquarks \cite{Agaev:2023wua}, whereas $X(6200)$
may be considered as a hadronic molecule $\eta _{c}\eta _{c}$ \cite%
{Agaev:2023ruu}. The structure $X(6900)$ can be interpreted as a
superposition of a diquark-antidiquark state built of pseudoscalar
components and a molecule $\chi _{c0}\chi _{c0}$ \cite%
{Agaev:2023ruu,Agaev:2023gaq}. In Ref.\ \cite{Agaev:2023rpj} we explained $%
X(7300)$ by employing the superposition of a molecule $\chi _{c1}\chi _{c1}$
and a radially excited diquark-antidiquark.

It is interesting that even in the framework of the four-quark picture there
are competing explanations for $X$ states. Thus, the resonance $X(6200)$ was
considered as the ground-state tetraquark with $J^{\mathrm{PC}}=0^{++}$ or $%
1^{+-}$. The first radially excited state of this tetraquark was assigned to
be $X(6600)$\ \cite{Wang:2022xja}. The $X$ resonances were interpreted as
different radially and orbitally excited diquark-antidiquark states also in
Refs. \cite{Dong:2022sef,Faustov:2022mvs}.

In most  articles devoted to analysis of fully charmed tetraquarks,
the authors investigated also their beauty partners $bb\overline{b}\overline{b}$
by computing the masses and other parameters of these particles. Such
structures, produced in  $pp$ and $p\overline{p}$ collisions, may be
discovered in the mass distributions of the $\eta _{b}\eta _{b}$, $\eta
_{b}\Upsilon $, and $\Upsilon \Upsilon $ mesons: In fact, $\Upsilon \Upsilon
$ pairs were detected and studied by CMS Collaboration \cite{CMS:2016liw}.

Predictions for parameters of $bb\overline{b}\overline{b}$ with different
quantum numbers were made in Refs.\ \cite%
{Berezhnoy:2011xn,Karliner:2016zzc,Esposito:2018cwh,Chen:2016jxd} using
various methods and schemes. Results of these articles sometimes contradict
to each another. For instance, in Ref.\ \cite{Berezhnoy:2011xn} it was shown
that the mass $18754~\mathrm{MeV}$ of the scalar exotic meson $X_{\mathrm{4b}%
}$ is below the $\eta _{b}\eta _{b}$ and $\Upsilon \Upsilon $ thresholds,
therefore it cannot be fixed in these mass distributions. A similar
problem was addressed in Ref.\ \cite{Karliner:2016zzc}, where the mass of $%
X_{\mathrm{4b}}$ was found equal to $(18826\pm 25)~\mathrm{MeV}$ which is
less than $\Upsilon \Upsilon $ but higher than $\eta _{b}\eta _{b}$ thresholds.
The masses of exotic $bb\overline{b}\overline{b}$ mesons with different
spin-parities were extracted from the sum rule analyses in Ref. \cite%
{Chen:2016jxd}. In accordance with this paper, the scalar tetraquarks have
masses $(18.45-18.59)\ \mathrm{GeV}$ and cannot be observed in
two-bottomonia final states. Only the scalar particle made of pseudoscalar
diquarks can decay to $\eta _{b}\eta _{b}$ and $\Upsilon \Upsilon $ mesons,
because its mass $(19640\pm 140)~\mathrm{MeV}$ considerably exceeds relevant
limits.

The scalar diquark-antidiquark states $X_{\mathrm{4b}}$ and $T_{\mathrm{4b}}$
with axial-vector and pseudoscalar diquarks were explored also in our
articles \cite{Agaev:2023wua,Agaev:2023gaq}. To this end, we used the QCD
two-point SR method \cite{Shifman:1978bx,Shifman:1978by}, which is one of
effective tools to investigate spectroscopic parameters and strong couplings
of conventional hadrons. But, it is also suitable to study multiquark
structures \cite{Albuquerque:2018jkn,Agaev:2020zad}. The masses $m=(18540\pm
50)~\mathrm{MeV}$ and $\widetilde{m}=(18858\pm 50)~\mathrm{MeV}$ of the
tetraquarks $X_{\mathrm{4b}}$ and $T_{\mathrm{4b}}$ found by this way,
allowed us to fix their possible strong decay modes. Because $m$ resides
below both $\eta _{b}\eta _{b}$ and $\Upsilon \Upsilon $ thresholds, this
particle cannot decay to two bottomonia final states. The exotic meson $T_{%
\mathrm{4b}}$ falls apart to a pair $\eta _{b}\eta _{b}$ and has the width
equal to $\widetilde{\Gamma }_{\mathrm{4b}}=(94\pm 28)~\mathrm{MeV}$ \cite%
{Agaev:2023gaq}.

But, fully beauty tetraquarks can also decay through alternative mechanisms
\cite{Karliner:2016zzc,Esposito:2018cwh,Becchi:2020mjz}. Thus, $X_{\mathrm{4b%
}}$ and $T_{\mathrm{4b}}$ can transform to $2\gamma $, $\Upsilon l^{+}l^{-}$%
or to four leptons $l_{1}^{+}l_{1}^{-}l_{2}^{+}l_{2}^{-}$ due to
annihilation of valence $b$ and $\overline{b}$ quarks and related processes.
The $b\overline{b}$ annihilations to gluons followed by appearance of
quark-antiquark pairs can generate processes with $\eta _{b}+H$, $B_{q}%
\overline{B}_{q}$, and $B_{q}^{\ast }\overline{B}_{q}^{\ast }\ (q=u,d,s,c)$
final states. It is clear that thresholds for these decays are considerably
smaller than masses of the tetraquarks $X_{\mathrm{4b}}$ and $T_{\mathrm{4b}%
} $. They are crucial for tetraquarks which are below the $\eta _{b}\eta
_{b} $ threshold and cannot dissociate to these bottomonia.

In the present work, we explore strong decays of the tetraquarks $X_{\mathrm{%
4b}}$ and $T_{\mathrm{4b}}$ to $B_{q}\overline{B}_{q}$ and $B_{q}^{\ast }%
\overline{B}_{q}^{\ast }$ mesons. In the case of $T_{\mathrm{4b}}$, they are
necessary to refine $\widetilde{\Gamma }_{\mathrm{4b}}$. But aforementioned
processes form a considerable part of $X_{\mathrm{4b}}$ tetraquark's full
width because a decay $X_{\mathrm{4b}}\rightarrow \eta _{b}\eta _{b}$ is
forbidden kinematically. Widths of decays under consideration are determined
by the strong couplings of particles at the vertices $X(T)_{\mathrm{4b}}B_{q}%
\overline{B}_{q}$ and $X(T)_{\mathrm{4b}}B_{q}^{\ast }\overline{B}_{q}^{\ast
}$. In the current article, we evaluate strong couplings of interest in the
context of the QCD three-point SR method.

This article is structured in the following manner: In Sec. \ref{sec:X4b},
we explore the decay channels of the tetraquark $X_{\mathrm{4b}}$ and
compute partial widths of the processes $X_{\mathrm{4b}}\rightarrow B_{q}%
\overline{B}_{q}$. The decays of $X_{\mathrm{4b}}$ to final states $B^{\ast
+}B^{\ast -}$, $\overline{B}^{\ast 0}B^{\ast 0}$, and $\overline{B}%
_{s}^{\ast 0}B_{s}^{\ast 0}$ are studied in Sec. \ref{sec:X4bVec}. Here, we
also evaluate the full width $X_{\mathrm{4b}}$. The similar investigation
for the diquark-antidiquark state $T_{\mathrm{4b}}$ is performed in Sec.\ %
\ref{sec:T4b}, in which we estimate contributions of the processes $T_{%
\mathrm{4b}}\rightarrow B_{q}\overline{B}_{q}$ and $T_{\mathrm{4b}%
}\rightarrow B_{q}^{\ast }\overline{B}_{q}^{\ast }$ to full width of $T_{%
\mathrm{4b}}$. In the last Section \ref{sec:Conc}, we compare obtained
predictions with available ones, and make our brief conclusions.

%%%%%%%%%%%%%%%%%%%%%%%%%%%%%%%%%%%%%%%%%%%%%%%%%%%%%%%%%%%%%%%%%%%%%%%%%%%%

\section{Decays $X_{\mathrm{4b}}\rightarrow B_{q}\overline{B}_{q}$}

\label{sec:X4b}

%%%%%%%%%%%%%%%%%%%%%%%%%%%%%%%%%%%%%%%%%%%%%%%%%%%%%%%%%%%

As we have noted above, $X_{\mathrm{4b}}$ cannot decay to meson pairs $\eta
_{b}\eta _{b}$ and $\Upsilon \Upsilon $, its full width is primarily
determined by the processes $X_{\mathrm{4b}}\rightarrow B_{q}\overline{B}%
_{q} $ and $X_{\mathrm{4b}}\rightarrow B_{q}^{\ast }\overline{B}_{q}^{\ast }$%
. Here, we evaluate the partial widths of the decays $X_{\mathrm{4b}%
}\rightarrow B^{+}B^{-}$, $\overline{B}^{0}B^{0}$, $\overline{B}%
_{s}^{0}B_{s}^{0}$, and $B_{c}^{+}B_{c}^{-}$, where $B_{(s,c)}$ are
pseudoscalar mesons.

The partial widths of these processes depend on the strong couplings $%
g_{l},\ l=1\div 4$ of the tetraquark $X_{\mathrm{4b}}$ and final state
mesons at the corresponding three-particle vertices. Therefore, the main
problem to be considered in this section is computation of $g_{l}$. In the
case of the channel $X_{\mathrm{4b}}\rightarrow B^{+}B^{-}$ this is a
coupling $g_{1}$ of particles at the vertex $X_{\mathrm{4b}}B^{+}B^{-}$. We
are going to analyze the decay $X_{\mathrm{4b}}\rightarrow B^{+}B^{-}$ in a
detailed manner and write down only essential expressions and numerical
results for other processes.

The strong coupling $g_{1}$ can be extracted from the three-point
correlation function%
\begin{eqnarray}
&&\Pi (p,p^{\prime })=i^{2}\int d^{4}xd^{4}ye^{ip^{\prime }y}e^{-ipx}\langle
0|\mathcal{T}\{J^{B^{+}}(y)  \notag \\
&&\times J^{B^{-}}(0)J^{\dagger }(x)\}|0\rangle ,  \label{eq:CF1}
\end{eqnarray}%
where $J(x)\ $is the interpolating current for the scalar tetraquark $X_{%
\mathrm{4b}}$%
\begin{equation}
J(x)=b_{a}^{T}(x)C\gamma _{\mu }b_{b}(x)\overline{b}_{a}(x)\gamma ^{\mu }C%
\overline{b}_{b}^{T}(x),  \label{eq:CRtetra}
\end{equation}%
with $C$ being the charge conjugation matrix.

The currents $J^{B^{+}}(x)$ and $J^{B^{-}}(x)$ for the $B$ mesons are given
by the formulas
\begin{equation}
J^{B^{+}}(x)=\overline{b}_{j}(x)i\gamma _{5}u_{j}(x),\ J^{B^{-}}(x)=%
\overline{u}_{i}(x)i\gamma _{5}b_{i}(x),  \label{eq:CRB}
\end{equation}%
where $i,\ j=1,2,3$ are color indices.

In accordance with the sum rule approach, we have to express the function $%
\Pi (p,p^{\prime })$ in terms of involved particles' parameters. By this
way, we determine the physical side of the sum rule. For these purposes, we
write down the $\Pi (p,p^{\prime })$ in the following form
\begin{eqnarray}
&&\Pi ^{\mathrm{Phys}}(p,p^{\prime })=\frac{\langle
0|J^{B^{+}}|B^{+}(p^{\prime })\rangle }{p^{\prime 2}-m_{B}^{2}}\frac{\langle
0|J^{B^{-}}|B^{-}(q)\rangle }{q^{2}-m_{B}^{2}}  \notag \\
&&\times \langle B^{+}(p^{\prime })B^{-}(q)|X_{\mathrm{4b}}(p)\rangle \frac{%
\langle X_{\mathrm{4b}}(p)|J^{\dagger }|0\rangle }{p^{2}-m^{2}}+\cdots ,
\label{eq:CF2}
\end{eqnarray}
where only the contribution of ground-level particles is presented
explicitly: Effects of higher resonances and continuum states are shown as
ellipses. It is evident, that four momenta of $X_{\mathrm{4b}}$ and $B^{+}$
are $p$ and $p^{\prime }$, respectively. Therefore, the momentum of $B^{-}$
is equal to $q=p-p^{\prime }$.

To simplify the correlation function $\Pi ^{\mathrm{Phys}}(p,p^{\prime })$,
we express the matrix elements which enter to Eq.\ (\ref{eq:CF2}), using the
masses and current couplings (decay constants) of involved particles. For
the scalar tetraquark $X_{\mathrm{4b}}$ the matrix element $\langle 0|J|X_{%
\mathrm{4b}}\rangle $ can be replaced by a product of its mass $m$ and
current coupling $f$
\begin{equation}
\langle 0|J|X_{\mathrm{4b}}\rangle =fm.  \label{eq:ME1}
\end{equation}%
The matrix element of the pseudoscalar $B$ mesons is determined by the
formula
\begin{equation}
\langle 0|J^{B}|B\rangle =\frac{f_{B}m_{B}^{2}}{m_{b}},  \label{eq:ME2}
\end{equation}%
with $m_{B}$ and $f_{B}$ being their mass and decay constant. Here, $m_{b}$
is the mass of the $b$-quark.

The vertex $\langle B^{+}(p^{\prime })B^{-}(q)|X_{\mathrm{4b}}(p)\rangle $
is modeled in the following form
\begin{equation}
\langle B^{+}(p^{\prime })B^{-}(q)|X_{\mathrm{4b}}(p)\rangle
=g_{1}(q^{2})p\cdot p^{\prime }.  \label{eq:ME3}
\end{equation}%
Here, $g_{1}(q^{2})$ is the form factor which at the mass shell of the $%
B^{-} $ meson, i.e., at $q^{2}=m_{B}^{2}$ fixes the strong coupling $g_{1}$.

By taking into account these expressions, it is not difficult to recast $\Pi
^{\mathrm{Phys}}(p,p^{\prime })$ into the form
\begin{eqnarray}
&&\Pi ^{\mathrm{Phys}}(p,p^{\prime })=g_{1}(q^{2})\frac{fmf_{B}^{2}m_{B}^{4}%
}{2m_{b}^{2}\left( p^{2}-m^{2}\right) \left( p^{\prime 2}-m_{B}^{2}\right) }
\notag \\
&&\times \frac{\left( m^{2}+m_{B}^{2}-q^{2}\right) }{(q^{2}-m_{B}^{2})}%
+\cdots ,  \label{eq:CorrF5}
\end{eqnarray}%
where the dots denote contributions of higher resonances and continuum
states. The correlator $\Pi ^{\mathrm{Phys}}(p,p^{\prime })$ is simply
proportional to $\mathrm{I}$. Therefore, the whole expression in the right
hand side of Eq.\ (\ref{eq:CorrF5}) is the invariant amplitude $\Pi ^{%
\mathrm{Phys}}(p^{2},p^{\prime 2},q^{2})$ which can be applied to derive the
form factor $g_{1}(q^{2})$.

The second component which is required to get the sum rule for $g_{1}(q^{2})$
is the correlator Eq.\ (\ref{eq:CF1}) computed using the quark propagators,
which reads%
\begin{eqnarray}
&&\Pi ^{\mathrm{OPE}}(p,p^{\prime })=\frac{16}{3}\int
d^{4}xd^{4}ye^{ip^{\prime }y}e^{-ipx}\langle \overline{b}b\rangle  \notag \\
&&\times \mathrm{Tr}\left[ S_{u}^{ij}(y)\gamma
_{5}S_{b}^{ja}(-x){}S_{b}^{ai}(x-y)\gamma _{5}\right] .  \label{eq:QCDside}
\end{eqnarray}%
In Eq.\ (\ref{eq:QCDside}) $S_{u(b)}(x)$ are $u$ and $b$ quark propagators%
\begin{eqnarray}
&&S_{u}^{ab}(x)=i\frac{\slashed x}{2\pi ^{2}x^{4}}\delta _{ab}-\frac{m_{u}}{%
4\pi ^{2}x^{2}}\delta _{ab}-\frac{\langle \overline{u}u\rangle }{12}\delta
_{ab}  \notag \\
&&+i\langle \overline{u}u\rangle \frac{m_{u}\slashed x}{48}\delta
_{ab}+\cdots ,  \label{eq:LQprop}
\end{eqnarray}%
and
\begin{eqnarray}
&&S_{b}^{ab}(x)=\frac{i}{(2\pi )^{4}}\int d^{4}ke^{-ikx}\Bigg \{\frac{\delta
_{ab}\left( {\slashed k}+m_{b}\right) }{k^{2}-m_{b}^{2}}  \notag \\
&&-\frac{g_{s}G_{ab}^{\alpha \beta }}{4}\frac{\sigma _{\alpha \beta }\left( {%
\slashed k}+m_{b}\right) +\left( {\slashed k}+m_{b}\right) \sigma _{\alpha
\beta }}{(k^{2}-m_{b}^{2})^{2}}  \notag \\
&&+\mathcal{O}\langle \alpha _{s}G^{2}/\pi \rangle \delta _{ab}+\cdots \Bigg
\}.  \label{eq:HQprop}
\end{eqnarray}%
Here, we have used the notation
\begin{equation}
G_{ab}^{\alpha \beta }\equiv G_{A}^{\alpha \beta }\lambda _{ab}^{A}/2,
\end{equation}%
where $G_{A}^{\alpha \beta }$ is the gluon field-strength tensor, and $%
\lambda ^{A}$ are the Gell-Mann matrices. The index $A$ runs in the range $%
1,2,\ldots 8$. The propagators $S_{u(b)}(x)$ are known with considerably
higher accuracy, but in Eqs.\ (\ref{eq:LQprop}) and (\ref{eq:HQprop}) we
keep only a few terms: Arguments in favor of such choices will be provided
below.

The correlator $\Pi ^{\mathrm{OPE}}(p,p^{\prime })$ contains three quark
propagators and vacuum condensate $\langle \overline{b}b\rangle $ of $b$
quarks. The function $\Pi ^{\mathrm{OPE}}(p,p^{\prime })$ differs from a
standard one which in the case, for instance, of the decay $T_{\mathrm{4b}%
}\rightarrow \eta _{b}\eta _{b}$ depends on four propagators $S_{b}(x)$. The
reason is that to calculate $\Pi ^{\mathrm{OPE}}(p,p^{\prime })$ one
contracts heavy and light quark fields, and because a pair of \ $B^{+}B^{-}$
mesons contains only $b$ and $\overline{b}$ quarks, remaining $\overline{b}b$
fields in $X_{\mathrm{4b}}$ constitute a heavy quark condensate.

Using the relation between the heavy quark and gluon condensates
\begin{equation}
m_{b}\langle \overline{b}b\rangle =-\frac{1}{12\pi }\langle \frac{\alpha
_{s}G^{2}}{\pi }\rangle ,  \label{eq:Conden}
\end{equation}%
we get
\begin{eqnarray}
&&\Pi ^{\mathrm{OPE}}(p,p^{\prime })=-\frac{4}{9m_{b}\pi }\langle \frac{%
\alpha _{s}G^{2}}{\pi }\rangle \int d^{4}xd^{4}ye^{ip^{\prime }y}e^{-ipx}
\notag \\
&&\times \mathrm{Tr}\left[ S_{u}^{ij}(y)\gamma
_{5}S_{b}^{ja}(-x){}S_{b}^{ai}(x-y)\gamma _{5}\right] .  \label{eq:CF6}
\end{eqnarray}%
In other words, the correlator $\Pi ^{\mathrm{OPE}}(p,p^{\prime })$ is
suppressed by the dimension-four factor $\langle \alpha _{s}G^{2}/\pi
\rangle $. In what follows, we denote by $\Pi ^{\mathrm{OPE}%
}(p^{2},p^{\prime 2},q^{2})$ the corresponding invariant amplitude.

In calculation of $\Pi ^{\mathrm{OPE}}(p,p^{\prime })$, we set $m_{u}=0$.
The perturbative terms in all propagators lead to a contribution which is
proportional to $\langle \alpha _{s}G^{2}/\pi \rangle $. A dimension-7 term
in $\Pi ^{\mathrm{OPE}}(p,p^{\prime })$ arising from the component $\sim
\langle \overline{u}u\rangle $ in $S_{u}(x)$ and perturbative ones in $%
S_{b}(x)$ vanishes. A contribution $\sim \langle \alpha _{s}G^{2}/\pi
\rangle ^{2}$ generated by components $g_{s}G_{ab}^{\alpha \beta }$ in $b$
propagators and perturbative term in $S_{u}(x)$ can be safely neglected.
Higher-dimensional pieces in the quark propagators omitted in Eqs.\ (\ref%
{eq:LQprop}) and (\ref{eq:HQprop}) give effects suppressed by additional
factors. As a result, $\Pi ^{\mathrm{OPE}}(p,p^{\prime })$ calculated with
dimension-7 accuracy actually contains a dimension-4 term proportional to $%
\langle \alpha _{s}G^{2}/\pi \rangle $.

To derive sum rule for the form factor $g_{1}(q^{2})$, we equate the
invariant amplitudes $\Pi ^{\mathrm{Phys}}(p^{2},p^{\prime 2},q^{2})$ and $%
\Pi ^{\mathrm{OPE}}(p^{2},p^{\prime 2},q^{2})$, and get the sum rule
equality. Contributions of the higher resonances and continuum terms can be
suppressed by applying Borel transformations over variables $-p^{2}$ and $%
-p^{\prime 2}$ to both sides of this expression, and removed using an
assumption on the quark-hadron duality. After these operations, we find
\begin{eqnarray}
&&g_{1}(q^{2})=\frac{2m_{b}^{2}}{fmf_{B}^{2}m_{B}^{4}}\frac{q^{2}-m_{B}^{2}}{%
m^{2}+m_{B}^{2}-q^{2}}e^{m^{2}/M_{1}^{2}}e^{m_{B}^{2}/M_{2}^{2}}  \notag \\
&&\times \Pi (\mathbf{M}^{2},\mathbf{s}_{0},q^{2}),  \label{eq:SRCoup}
\end{eqnarray}%
where $\Pi (\mathbf{M}^{2},\mathbf{s}_{0},q^{2})$ is the amplitude $\Pi ^{%
\mathrm{OPE}}(p^{2},p^{\prime 2},q^{2})$ after Borel transformations and
continuum subtractions. It can be expressed by means of the spectral density
$\rho (s,s^{\prime },q^{2})$
\begin{eqnarray}
&&\Pi (\mathbf{M}^{2},\mathbf{s}_{0},q^{2})=\int_{16m_{b}^{2}}^{s_{0}}ds%
\int_{m_{b}^{2}}^{s_{0}^{\prime }}ds^{\prime }\rho (s,s^{\prime },q^{2})
\notag \\
&&\times e^{-s/M_{1}^{2}}e^{-s^{\prime }/M_{2}^{2}}.  \label{eq:SCoupl}
\end{eqnarray}%
Here, $(M_{1}^{2},s_{0})$ and $(M_{2}^{2},s_{0}^{\prime })$ are the Borel
and continuum subtraction parameters for the $X_{\mathrm{4b}}$ and $B^{+}$
channels, respectively. It is worth noting that $\rho (s,s^{\prime },q^{2})$
is computed as an imaginary part of the correlation function $\Pi ^{\mathrm{%
OPE}}(p,p^{\prime })$.

As is seen, $g_{1}(q^{2})$ contains the mass $m$ and current coupling $f$ of
the tetraquark $X_{\mathrm{4b}}$. These quantities were found in Ref. \cite%
{Agaev:2023wua}
\begin{eqnarray}
m &=&(18540\pm 50)~\mathrm{MeV},  \notag \\
f &=&(6.1\pm 0.4)\times 10^{-1}~\mathrm{GeV}^{4}.  \label{eq:Result1}
\end{eqnarray}%
To this end, we used the two-point SR method and applied for the Borel and
continuum subtraction parameters the following regions
\begin{equation}
M^{2}\in \lbrack 17.5,18.5]~\mathrm{GeV}^{2},\ s_{0}\in \lbrack 375,380]~%
\mathrm{GeV}^{2}.  \label{eq:Wind1}
\end{equation}

The sum rule Eq.\ (\ref{eq:SRCoup}) depends also on the mass $%
m_{B}=(5279.25\pm 0.26)~\mathrm{MeV}$ and decay constant $f_{B}=(206\pm 7)~%
\mathrm{MeV}$ of the $B^{\pm }$ mesons borrowed from Refs.\ \cite{PDG:2022}
and \cite{Narison:2012xy}, respectively. The values of the gluon condensate,
and $b$ and $c$ quarks' masses are well known
\begin{eqnarray}
&&\langle \frac{\alpha _{s}G^{2}}{\pi }\rangle =(0.012\pm 0.004)~\mathrm{GeV}%
^{4},  \notag \\
&&m_{b}=4.18_{-0.02}^{+0.03}~\mathrm{GeV},  \notag \\
&&m_{c}=(1.27\pm 0.02)~\mathrm{GeV}.  \label{Parameters}
\end{eqnarray}

To perform numerical analysis, one has to fix the working windows for the
parameters $(M_{1}^{2},s_{0})$ and $(M_{2}^{2},s_{0}^{\prime })$. For $%
M_{1}^{2}$ and $s_{0}$ connected with the tetraquark $X_{\mathrm{4b}}$, we
employ the regions Eq.\ (\ref{eq:Wind1}). It is worth noting that the
working windows in Eq.\ (\ref{eq:Wind1}) meet all constraints imposed on them
by SR method. Thus, the pole contribution $\mathrm{PC}$ in the relevant mass
calculations changes within limits $0.72\geq \mathrm{PC}\geq 0.66$. In other
words, for all $s_{0}$ the pole contribution exceeds $0.5$. At $%
M_{1}^{2}=17.5~\mathrm{GeV}^{2}$ a dimension-$4$ term constitutes $\simeq
-1.5\%$ of the result ensuring convergence of the operator product
expansion. Such strong constraints naturally lead to rather narrow regions
for $M_{1}^{2}$ and $s_{0}$. Because $g_{1}(q^{2})$ depends also on $fm$,
another choice for $(M_{1}^{2},s_{0})$ may generate additional,
uncontrollable ambiguities.

The parameters $(M_{2}^{2},\ s_{0}^{\prime })$ for the $B^{+}$ channel are
chosen within limits
\begin{equation}
M_{2}^{2}\in \lbrack 5.5,6.5]~\mathrm{GeV}^{2},\ s_{0}^{\prime }\in \lbrack
33.5,34.5]~\mathrm{GeV}^{2}.  \label{eq:Wind2}
\end{equation}%
The $s_{0}^{\prime }$ is limited by the mass $m_{B^{+}(2S)}=5976~\mathrm{MeV}
$ of the excited $B^{+}(2S)$ meson \cite{Ebert:2009ua} and satisfies $%
s_{0}^{\prime }<m_{B^{+}(2S)}^{2}$. The Borel parameter $M_{2}^{2}$ also
complies with constraints of SR analysis. But these two sets $%
(M_{1}^{2},s_{0})$ and $(M_{2}^{2},s_{0}^{\prime })$ should lead to
relatively stable regions where $g_{1}(q^{2})$ can be evaluated.

\begin{figure}[h]
\includegraphics[width=8.5cm]{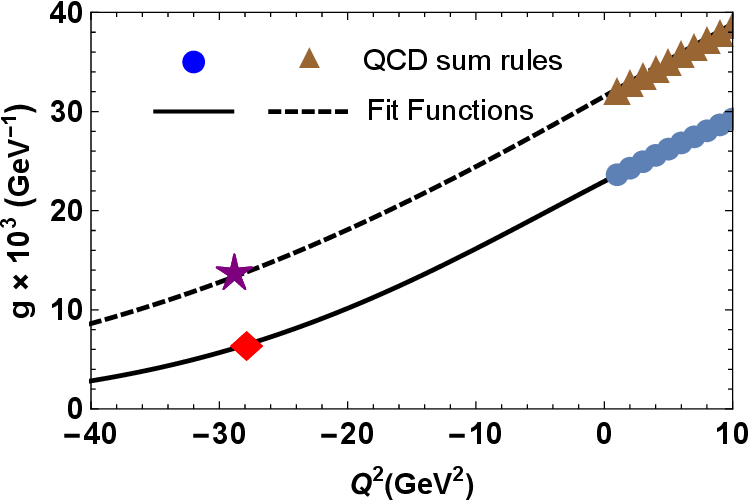}
\caption{QCD predictions and $\mathcal{F}(Q^2)$ functions for the form
factors $g_{1}(Q^{2})$ (solid line) and $g_{5}(Q^{2})$ (dashed line). The
couplings $g_{1}$ and $g_{5}$ are extracted at the points $Q^{2}=-m_{B}^{2}$
and $Q^{2}=-m_{B^{*}}^{2}$ labeled on the plot by the red diamond and star,
respectively.}
\label{fig:Fit}
\end{figure}

\begin{figure}[h]
\includegraphics[width=8.8cm]{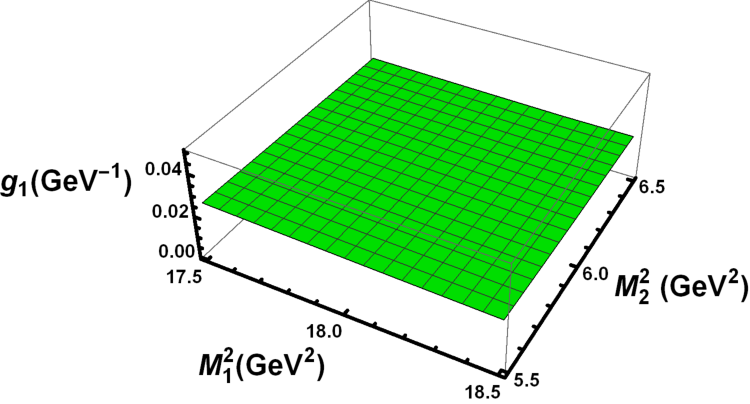}
\caption{ The strong coupling $g=g_{1}(2~\mathrm{GeV}^{2})$ as a function of
the parameters $M_{1}^{2}$ and $M_{2}^{2}$ at $s_{0}=377.5~\mathrm{GeV}^{2}$
and $s_{0}^{\prime }=34~\mathrm{GeV}^{2}$.}
\label{fig:StrongC}
\end{figure}

It is known that the sum rule method gives credible results only in the
Euclidean region $q^{2}<0$. Therefore, we introduce a new variable $%
Q^{2}=-q^{2}$ and denote the obtained function by $g_{1}(Q^{2})$. We compute
$g_{1}(Q^{2})$ by varying $Q^{2}$ within the boundaries $Q^{2}=1\div 10~%
\mathrm{GeV}^{2}$ and depict obtained results in Fig.\ \ref{fig:Fit}. Let us
emphasize that at each $Q^{2}$ calculations performed here meet constraints
imposed on parameters $\mathbf{M}^{2}$ and $\mathbf{s}_{0}$ by SR method.
For example, in Fig.\ \ref{fig:StrongC} the coupling $g_{1}(2~\mathrm{GeV}%
^{2})$ is plotted as a function of the parameters $M_{1}^{2}$ and $M_{2}^{2}$
at middle of the regions $s_{0}$ and $s_{0}^{\prime }$, where it
demonstrates a relative stability: Indeed, upon changing $M_{1}^{2}$ and $%
M_{2}^{2}$ inside of explored regions, variations of $g_{1}(2~\mathrm{GeV}%
^{2})$ do not exceed $\pm 12\%$ of the central value. At the point $Q^{2}=2~%
\mathrm{GeV}^{2}$, we get
\begin{equation}
g_{1}(2~\mathrm{GeV}^{2})=(2.30\pm 0.26)\times 10^{-2}\ \mathrm{GeV}^{-1}.
\end{equation}

To calculate the partial width of the process $X_{\mathrm{4b}}\rightarrow
B^{+}B^{-}$, one needs the value of the form factor $g_{1}(q^{2})$ at the
mass shell of the $B^{-}$ meson $q^{2}=m_{B}^{2}$. To this end, it is
necessary to introduce a fit function $\mathcal{F}_{1}(Q^{2})$ that at
momenta $Q^{2}>0$ leads to the same data as the SR computations, but can be
extrapolated to a region of $Q^{2}<0$ and employed to fix $\mathcal{F}%
_{1}(-m_{B}^{2})$.

In present article, we use the functions $\mathcal{F}_{l}(Q^{2})$
\begin{equation}
\mathcal{F}_{l}(Q^{2})=\mathcal{F}_{l}^{0}\mathrm{\exp }\left[ c_{l}^{1}%
\frac{Q^{2}}{m^{2}}+c_{l}^{2}\left( \frac{Q^{2}}{m^{2}}\right) ^{2}\right] ,
\label{eq:FitF}
\end{equation}%
with parameters $\mathcal{F}_{l}^{0}$, $c_{l}^{1}$ and $c_{l}^{2}$. They
should be fixed by comparing SR predictions and $\mathcal{F}_{1}(Q^{2})$.
Analysis carried out in the case of the form factor $g_{1}(q^{2})$ gives $%
\mathcal{F}_{1}^{0}=0.02\ \mathrm{GeV}^{-1}$, $c_{1}^{1}=10.07$, and $%
c_{1}^{2}=-68.03$. This function is depicted in Fig.\ \ref{fig:Fit}, where
one can be convinced in nice agreement of $\mathcal{F}_{1}(Q^{2})$ and QCD
SR data.

For the coupling $g_{1}$, we find
\begin{equation}
g_{1}\equiv \mathcal{F}_{1}(-m_{B}^{2})=(6.49\pm 1.41)\times 10^{-3}\
\mathrm{GeV}^{-1},  \label{eq:Coupl1}
\end{equation}%
where ambiguities in Eq.\ (\ref{eq:ME4}) are generated mainly by the choice
of the parameters $\mathbf{M}^{2}$ and $\mathbf{s}_{0}$. Let us note that,
we calculate the correlation function $\Pi ^{\mathrm{OPE}}(p,p^{\prime })$
and $g_{1}$ (and other strong couplings) at leading order of QCD. In
general, the next-to-leading order (NLO) perturbative contributions improve
accuracy of theoretical analysis and are necessary to fix a scale $\mu $ in
heavy quark masses and vacuum condensates. Depending on the  problem under
consideration, NLO terms may affect the final results. Indeed, NLO corrections
to parameters of light four-quark mesons are significant \cite{Ray:2022fcl}.
At the same time, similar contributions to masses of doubly heavy
tetraquarks calculated in Ref.\ \cite{Albuquerque:2022weq} by means of the
inverse Laplace SR approach were found numerically small. The smallness of NLO corrections in mass computations may be explained by an analytic form of the relevant SR given as a ratio of two-point correlation functions. This is not the case for vertex functions, where NLO effects may be large. But this problem requires detailed studies, which are beyond the scope of our paper.

The width of the channel $X_{\mathrm{4b}}\rightarrow B^{+}B^{-}$ is given by
the formula%
\begin{equation}
\Gamma \left[ X_{\mathrm{4b}}\rightarrow B^{+}B^{-}\right] =g_{1}^{2}\frac{%
m_{B}^{2}\lambda }{8\pi }\left( 1+\frac{\lambda ^{2}}{m_{B}^{2}}\right) ,
\label{eq:PDw2}
\end{equation}%
where $\lambda =\lambda (m,m_{B},m_{B})$, and
\begin{equation}
\lambda (x,y,z)=\frac{\sqrt{%
x^{4}+y^{4}+z^{4}-2(x^{2}y^{2}+x^{2}z^{2}+y^{2}z^{2})}}{2x}.
\end{equation}%
We find
\begin{equation}
\Gamma _{1}\left[ X_{\mathrm{4b}}\rightarrow B^{+}B^{-}\right] =(1.10\pm
0.34)~\mathrm{MeV}.  \label{eq:DW2}
\end{equation}

The parameters of the decay $X_{\mathrm{4b}}\rightarrow \overline{B}%
^{0}B^{0} $, almost coincide with ones for the process $X_{\mathrm{4b}%
}\rightarrow B^{+}B^{-}$ though there is a small gap between the masses $%
m_{B_{0}}=(5279.63\pm 0.20)~\mathrm{MeV}$ and $m_{B}$ of the mesons $B^{0}$
and $B^{\pm }$. As a result, one obtains $g_{2}(q^{2})\approx g_{1}(q^{2})$
and $\Gamma _{2}\left[ X_{\mathrm{4b}}\rightarrow \overline{B}^{0}B^{0}%
\right] \approx \Gamma _{1}\left[ X_{\mathrm{4b}}\rightarrow B^{+}B^{-}%
\right] $.

The analysis of the process $X_{\mathrm{4b}}\rightarrow \overline{B}%
_{s}^{0}B_{s}^{0}$ requires some modifications. First of all, the
interpolating currents of the mesons $\overline{B}_{s}^{0}$ and $B_{s}^{0}$
are
\begin{equation}
J^{\overline{B}_{s}}(x)=\overline{s}_{j}(x)i\gamma _{5}b_{j}(x),\
J^{B_{s}}(x)=\overline{b}_{i}(x)i\gamma _{5}s_{i}(x).
\end{equation}%
The matrix element of $\overline{B}_{s}^{0}$ and $B_{s}^{0}$ mesons is
\begin{equation}
\langle 0|J^{B_{s}}|B_{s}^{0}\rangle =\frac{f_{B_{s}}m_{B_{s}}^{2}}{%
m_{b}+m_{s}},  \label{eq:ME4}
\end{equation}%
with $m_{s}=93.4_{-3.4}^{+8.6}~\mathrm{MeV}$ being the $s$ quark's mass. In
Eq.\ (\ref{eq:ME4}) $m_{B_{s}}=(5366.91\pm 0.11)~\mathrm{MeV}$ and $%
f_{B_{s}}=(234\pm 5)~\mathrm{MeV}$ are the mass and decay constant of these
mesons. Therefore, Eqs.\ (\ref{eq:SRCoup}) and (\ref{eq:SCoupl}) change
accordingly, where one should replace $m_{b}^{2}\rightarrow
(m_{b}+m_{s})^{2} $.

In computations of $g_{3}(q^{2})$, we use the following Borel and continuum
subtraction parameters: for the $X_{\mathrm{4b}}$ channel the parameters $%
(M_{1}^{2},s_{0})$ from Eq.\ (\ref{eq:Wind1}), and
\begin{equation}
M_{2}^{2}\in \lbrack 5.5,6.5]~\mathrm{GeV}^{2},\ s_{0}^{\prime }\in \lbrack
34,35]~\mathrm{GeV}^{2},  \label{eq:Wind3}
\end{equation}%
for the $\overline{B}_{s}^{0}$ channel. The fit function $\mathcal{F}%
_{3}(Q^{2})$ necessary for further computations is determined by the
parameters $\mathcal{F}_{3}^{0}=0.02\ \mathrm{GeV}^{-1}$, $c_{3}^{1}=9.67$,
and $c_{3}^{2}=-63.84$. Then for the coupling $g_{3}$ and width of the decay
$X_{\mathrm{4b}}\rightarrow \overline{B}_{s}^{0}B_{s}^{0}$, we find
\begin{eqnarray}
&&g_{3}\equiv \mathcal{F}_{3}(-m_{B_{s}}^{2})=(5.61\pm 1.21)\times 10^{-3}\
\mathrm{GeV}^{-1},  \notag \\
&&\Gamma _{3}\left[ X_{\mathrm{4b}}\rightarrow \overline{B}_{s}^{0}B_{s}^{0}%
\right] =(0.81\pm 0.25)~\mathrm{MeV}.  \label{eq:Results3}
\end{eqnarray}

The investigations of the fourth decay $X_{\mathrm{4b}}\rightarrow
B_{c}^{+}B_{c}^{-}$ can be carried out in the standard way. In this case, we
consider the correlation function, in which the interpolating currents for
the mesons $B_{c}^{+}$ and $B_{c}^{-}$ have the forms
\begin{equation}
J^{B_{c}^{+}}(x)=\overline{b}_{j}(x)i\gamma _{5}c_{j}(x),\ J^{B_{c}^{-}}(x)=%
\overline{c}_{i}(x)i\gamma _{5}b_{i}(x).
\end{equation}%
The matrix element of these mesons is%
\begin{equation}
\langle 0|J^{B_{c}}|B_{c}\rangle =\frac{f_{B_{c}}m_{B_{c}}^{2}}{m_{b}+m_{c}},
\label{eq:ME5}
\end{equation}%
with $m_{B_{c}}=(6274.47\pm 0.27)~\mathrm{MeV}$ and $f_{B_{c}}=(476\pm 27)~%
\mathrm{MeV}$ being the mass and decay constant of $B_{c}^{\pm }$ \cite%
{PDG:2022,Veliev:2010vd}. The form factor $g_{4}(q^{2})$ is extracted from
the SRs using the following parameters
\begin{equation}
M_{2}^{2}\in \lbrack 6.5,7.5]~\mathrm{GeV}^{2},\ s_{0}^{\prime }\in \lbrack
45,47]~\mathrm{GeV}^{2}.  \label{eq:Wind3A}
\end{equation}%
The fit function $\mathcal{F}_{4}(Q^{2})$ is given by the parameters $%
\mathcal{F}_{4}^{0}=0.009\ \mathrm{GeV}^{-1}$, $c_{4}^{1}=4.58$, and $%
c_{4}^{2}=-8.89$. Then for the coupling $g_{4}$ and width of the decay $X_{%
\mathrm{4b}}\rightarrow B_{c}^{+}B_{c}^{-}$, we get
\begin{eqnarray}
&&g_{4}\equiv \mathcal{F}_{4}(-m_{B_{c}}^{2})=(4.67\pm 0.89)\times 10^{-3}\
\mathrm{GeV}^{-1},  \notag \\
&&\Gamma _{4}\left[ X_{\mathrm{4b}}\rightarrow B_{c}^{+}B_{c}^{-}\right]
=(0.51\pm 0.14)~\mathrm{MeV}.  \label{eq:Results3A}
\end{eqnarray}%
Predictions obtained for parameters of these decays are collected in Table\ %
\ref{tab:Channels}.

\begin{widetext}

\begin{table}[tbp]
\begin{tabular}{|c|c|c|c|c|c|c|c|}
\hline\hline
$l$ & Modes &  $M_2^2 (\mathrm{GeV}^2)$ & $s_{0}^{\prime} (\mathrm{GeV}^2)$ & $g_{l} ~(\mathrm{GeV}^{-1})$ & $\Gamma_{l}(\mathrm{MeV})$ & $G_{l} ~(\mathrm{GeV}^{-1})$ & $\widetilde{\Gamma}_{l}(\mathrm{MeV})$  \\ \hline
$1$ & $ B^{+}B^{-}$ &  $5.5-6.5$ & $33.5-34.5$ & $(6.49 \pm 1.41)\times 10^{-3}$ & $1.10 \pm 0.34$ & $(1.02 \pm 0.21)\times 10^{-2}$ & $2.92 \pm 0.89$  \\
$2$ & $\overline{B}^{0}B^{0}$ &  $5.5-6.5$ & $33.5-34.5$ &$(6.49 \pm 1.41)\times 10^{-3}$ & $1.10 \pm 0.34$ & $(1.02 \pm 0.21)\times 10^{-2}$ & $2.92 \pm 0.89$ \\
$3$ & $\overline{B}_{s}^{0}B_{s}^{0}$ &  $5.5-6.5$ & $34-35$ & $(5.61 \pm 1.21)\times 10^{-3}$ & $0.81 \pm 0.25 $& $(0.85 \pm 0.17)\times 10^{-2}$ & $1.99 \pm 0.59$  \\
$4$ & $ B_{c}^{+}B_{c}^{-}$ & $6.5-7.5$ & $45-47$ & $(4.67 \pm 0.89)\times 10^{-3}$ & $0.51 \pm 0.14$ & $(4.87 \pm 0.95)\times 10^{-3}$ & $0.59 \pm 0.16$  \\
$5$ &  $ B^{\ast +}B^{\ast -}$ &  $5.5-6.5$ & $34-35$ & $(1.36 \pm 0.26)\times 10^{-2}$ & $2.26\pm 0.62$ & $(1.75 \pm 0.35)\times 10^{-2}$ & $4.05 \pm 1.15$  \\
$6$ & $\overline{B}^{\ast 0}B^{\ast 0}$ &  $5.5-6.5$ & $34-35$ & $(1.36 \pm 0.26)\times 10^{-2}$ & $2.26\pm 0.62$ & $(1.75 \pm 0.35)\times 10^{-2}$ & $4.05 \pm 1.15$  \\
$7$ & $\overline{B}_{s}^{\ast 0}B_{s}^{\ast 0}$ & $6-7$ & $35-36$ & $(1.14 \pm 0.22)\times 10^{-2}$ & $1.58 \pm 0.43$ & $(1.59 \pm 0.32)\times 10^{-2}$ & $3.32 \pm 0.95$ \\
 \hline\hline
\end{tabular}%
\caption{Decay modes of the tetraquarks $X_{\mathrm{4b}}$ and $T_{\mathrm{4b}}$,
strong couplings $g_{l}$, $G_{l}$, and partial widths $\Gamma_{l}$, $\widetilde{\Gamma}_{l}$. For all decays the Borel and continuum subtraction parameters in $X_{\mathrm{4b}}$ channel are $M_{1}^{2}\in \lbrack 17.5,18.5]~\mathrm{GeV}^{2}$ and $s_{0}\in \lbrack 375,380]~\mathrm{GeV}^{2}$, whereas in $T_{\mathrm{4b}}$ channel  $\widetilde{M}_{1}^{2}\in \lbrack 17.5,18.5]~\mathrm{GeV}^{2}$ and $\widetilde{s}_{0}\in \lbrack 380,385]~\mathrm{GeV}^{2}$ have been used.  }
\label{tab:Channels}
\end{table}

\end{widetext}

%%%%%%%%%%%%%%%%%%%%%%%%%%%%%%%%%%%%%%%%%%%%%%%%%%%%%%%%%%%%%%%%%%%%%%%%%%%%%%%%%

\section{Channels $X_{\mathrm{4b}}\rightarrow B^{\ast +}B^{\ast -}$, $%
\overline{B}^{\ast 0}B^{\ast 0}$, and $\overline{B}_{s}^{\ast 0}B_{s}^{\ast
0}$}

\label{sec:X4bVec}

%%%%%%%%%%%%%%%%%%%%%%%%%%%%%%%%%%%%%%%%%%%%%%%%%%%%%%%%%%%%%%%%%%%%%%%%%%%%

The decays of the tetraquark $X_{\mathrm{4b}}$ to vector mesons $B_{q}^{\ast
}\overline{B}_{q}^{\ast }$, with some modifications, can be analyzed as ones
studied in the previous section. Let us consider the process $X_{\mathrm{4b}%
}\rightarrow B^{\ast +}B^{\ast -}$ and evaluate the form factor $%
g_{5}(q^{2}) $ corresponding to the vertex $X_{\mathrm{4b}}B^{\ast +}B^{\ast
-}$.

The correlation function required to derive SR for $g_{5}(q^{2})$ is%
\begin{eqnarray}
&&\Pi _{\mu \nu }(p,p^{\prime })=i^{2}\int d^{4}xd^{4}ye^{ip^{\prime
}y}e^{-ipx}\langle 0|\mathcal{T}\{J_{\mu }^{B^{\ast +}}(y)  \notag \\
&&\times J_{\nu }^{B^{\ast -}}(0)J^{\dagger }(x)\}|0\rangle ,  \label{eq:CF3}
\end{eqnarray}%
where
\begin{equation}
J_{\mu }^{B^{\ast +}}(x)=\overline{b}_{j}(x)\gamma _{\mu }u_{j}(x),\ J_{\nu
}^{B^{\ast -}}(x)=\overline{u}_{i}(x)\gamma _{\nu }b_{i}(x)  \label{eq:Curr1}
\end{equation}%
are interpolating currents for the vector mesons $B^{\ast +}$ and $B^{\ast
-} $, respectively.

The $\Pi _{\mu \nu }(p,p^{\prime })$ in terms of physical parameters of the
particles $X_{\mathrm{4b}}$, $B^{\ast +}$, and $B^{\ast -}$ has the
decomposition%
\begin{eqnarray}
&&\Pi _{\mu \nu }^{\mathrm{Phys}}(p,q)=\frac{g_{5}(q^{2})fmf_{B^{\ast
}}^{2}m_{B^{\ast }}^{2}}{\left( p^{2}-m^{2}\right) \left( p^{\prime
2}-m_{B^{\ast }}^{2}\right) \left( q^{2}-m_{B^{\ast }}^{2}\right) }  \notag
\\
&&\times \left( \frac{m^{2}-m_{B^{\ast }}^{2}-q^{2}}{2}g_{\mu \nu }-p_{\mu
}p_{\nu }^{\prime }\right) +\cdots .  \label{eq:PhysSide}
\end{eqnarray}%
The Eq.\ (\ref{eq:PhysSide}) is obtained using the matrix elements
\begin{eqnarray}
\langle 0|J_{\mu }^{B^{\ast +}}|B^{\ast +}(p^{\prime })\rangle &=&f_{B^{\ast
}}m_{B^{\ast }}\varepsilon _{\mu }(p^{\prime }),  \notag \\
\langle 0|J_{\nu }^{B^{\ast -}}|B^{\ast -}(q)\rangle &=&f_{B^{\ast
}}m_{B^{\ast }}\varepsilon _{\nu }^{\prime }(q),
\end{eqnarray}%
where $m_{B^{\ast }}=(5324.71\pm 0.21)~\mathrm{MeV}$ and $f_{B^{\ast
}}=(210\pm 6)~\mathrm{MeV}$ are the mass and decay constants of the $B^{\ast
\pm }$ mesons \cite{PDG:2022,Narison:2015nxh}. Here, $\varepsilon _{\mu
}(p^{\prime })$ and $\varepsilon _{\nu }^{\prime }(q)$ are the polarization
vectors of $B^{\ast +}$ and $B^{\ast -}$, respectively. The vertex $X_{%
\mathrm{4b}}B^{\ast +}B^{\ast -}$ is modeled by the expression

\begin{eqnarray}
&&\langle B^{\ast +}\left( p^{\prime }\right) B^{\ast -}(q)|X_{\mathrm{4b}%
}(p)\rangle =g_{5}(q^{2})\left[ \left( q\cdot p^{\prime }\right) \right.
\notag \\
&&\left. \times \varepsilon ^{\ast }(p^{\prime })\cdot \varepsilon ^{\prime
\ast }(q)-p^{\prime }\cdot \varepsilon ^{\prime \ast }(q)q\cdot \varepsilon
^{\ast }(p^{\prime })\right] ,  \label{eq:Mel2}
\end{eqnarray}

The QCD side of SRs is determined by the formula
\begin{eqnarray}
&&\Pi _{\mu \nu }^{\mathrm{OPE}}(p,p^{\prime })=-\frac{4}{9m_{b}\pi }\langle
\frac{\alpha _{s}G^{2}}{\pi }\rangle \int d^{4}xd^{4}ye^{ip^{\prime
}y}e^{-ipx}  \notag \\
&&\times \mathrm{Tr}\left[ S_{u}^{ij}(y)\gamma _{\nu
}S_{b}^{ja}(-x){}S_{b}^{ai}(x-y)\gamma _{\mu }\right] .  \label{eq:CF7}
\end{eqnarray}

In the SR computations, we make use of invariant amplitudes $\widehat{\Pi }^{%
\mathrm{Phys}}(p^{2},p^{\prime 2},q^{2})$ and $\widehat{\Pi }^{\mathrm{OPE}%
}(p^{2},p^{\prime 2},q^{2})$ which correspond to terms $g_{\mu \nu }$ in
the physical and QCD sides, respectively. After Borel transformations and
continuum subtractions the SR for the form factor $g_{5}(q^{2})$ reads
\begin{eqnarray}
&&g_{5}(q^{2})=\frac{2}{fmf_{B^{\ast }}^{2}m_{B^{\ast }}^{2}}\frac{%
q^{2}-m_{B^{\ast }}^{2}}{m^{2}-m_{B^{\ast }}^{2}-q^{2}}e^{m^{2}/M_{1}^{2}}
\notag \\
&&\times e^{m_{B^{\ast }}^{2}/M_{2}^{2}}\widehat{\Pi }(\mathbf{M}^{2},%
\mathbf{s}_{0},q^{2}),
\end{eqnarray}%
with $\widehat{\Pi }(\mathbf{M}^{2},\mathbf{s}_{0},q^{2})$ being the
amplitude $\widehat{\Pi }^{\mathrm{OPE}}(p^{2},p^{\prime 2},q^{2})$ after
relevant manipulations.

The partial width of the decay $X_{\mathrm{4b}}\rightarrow B^{\ast +}B^{\ast
-}$ can be evaluated by means of the expression
\begin{equation}
\Gamma _{5}\left[ X_{\mathrm{4b}}\rightarrow B^{\ast +}B^{\ast -}\right]
=g_{5}^{2}\frac{m_{B^{\ast }}^{2}\widehat{\lambda }}{8\pi }\left( \frac{%
m_{B^{\ast }}^{2}}{m^{2}}+\frac{2\widehat{\lambda }^{2}}{3m_{B^{\ast }}^{2}}%
\right) ,
\end{equation}%
where $\widehat{\lambda }=\lambda (m,m_{B^{\ast }},m_{B^{\ast }})$.

The coupling $g_{5}$ is determined in accordance with a scheme explained
above. In Fig.\ \ref{fig:Fit}, we provide the SR data and extrapolating
function $\mathcal{F}_{5}(Q^{2})$ employed to find $g_{5}$. To extract the
SR data, we have used the working regions Eq.\ (\ref{eq:Wind3}). The
coupling $g_{5}$ has been evaluated at the mass shell of $B^{\ast -}$ meson.
This coupling and partial width of the decay $X_{\mathrm{4b}}\rightarrow
B^{\ast +}B^{\ast -}$ are equal to
\begin{eqnarray}
&&g_{5}\equiv \mathcal{F}_{5}(-m_{B^{\ast }}^{2})=(1.36\pm 0.26)\times
10^{-3}\ \mathrm{GeV}^{-1},  \notag \\
&&\Gamma _{5}\left[ X_{\mathrm{4b}}\rightarrow B^{\ast +}B^{\ast -}\right]
=(2.26\pm 0.62)~\mathrm{MeV}.
\end{eqnarray}

Predictions obtained for parameters of other modes are presented in Table\ %
\ref{tab:Channels}. Here, one can find couplings $g_{6}$ and $g_{7}$, as
well as partial widths of the decays $X_{\mathrm{4b}}\rightarrow \overline{B}%
^{\ast 0}B^{\ast 0}$ and $\overline{B}_{s}^{\ast 0}B_{s}^{\ast 0}$. The
strong coupling $g_{6}$ of particles at the vertex $X_{\mathrm{4b}}\overline{%
B}^{\ast 0}B^{\ast 0}$ and width of the channel $X_{\mathrm{4b}}\rightarrow
\overline{B}^{\ast 0}B^{\ast 0}$ do not differ numerically from those for
the process $X_{\mathrm{4b}}\rightarrow B^{\ast +}B^{\ast -}$. To calculate
parameters of the mode $X_{\mathrm{4b}}\rightarrow \overline{B}_{s}^{\ast
0}B_{s}^{\ast 0}$, we have used the following input information: the mass of
the $B_{s}^{\ast 0}$ meson $m_{B_{s}^{\ast }}=(5415.8\pm 1.5)~\mathrm{MeV}\ $
and its decay constant $f_{B_{s}^{\ast }}=(221\pm 7)~\mathrm{MeV}$. \ In
this case the regions for $M_{2}^{2}$ and $s_{0}^{\prime }$ are chosen as
\begin{equation}
M_{2}^{2}\in \lbrack 6,7]~\mathrm{GeV}^{2},\ s_{0}^{\prime }\in \lbrack
35,36]~\mathrm{GeV}^{2}.
\end{equation}%
For the width of the decay $X_{\mathrm{4b}}\rightarrow \overline{B}%
_{s}^{\ast 0}B_{s}^{\ast 0}$, we obtain%
\begin{equation}
\Gamma _{7}\left[ X_{\mathrm{4b}}\rightarrow \overline{B}_{s}^{\ast
0}B_{s}^{\ast 0}\right] =(1.58\pm 0.43)~\mathrm{MeV}.
\end{equation}

Information about partial widths of the decays obtained in last two
sections, allows us to estimate the full width of the tetraquark $X_{\mathrm{%
4b}}$%
\begin{equation}
\Gamma _{\mathrm{4b}}=(9.62\pm 1.13)~\mathrm{MeV}.
\end{equation}

%%%%%%%%%%%%%%%%%%%%%%%%%%%%%%%%%%%%%%%%%%%%%%%%%%%%%%%%%%%%%%%%%%%%%%%%%%%%

\section{Processes $T_{\mathrm{4b}}\rightarrow B_{q}\overline{B}_{q}$ and $%
T_{\mathrm{4b}}\rightarrow B_{q}^{\ast }\overline{B}_{q}^{\ast }$}

\label{sec:T4b} %%%%%%%%%%%%%%%%%%%%%%%%%%%%%%%%%%%%%%%%%%%%%%%%%%%%%%%%%%%

The scalar tetraquark $T_{\mathrm{4b}}$ was studied in Ref. \cite%
{Agaev:2023gaq}, in which we calculated the mass $\widetilde{m}$ and full
width $\widetilde{\Gamma }_{\mathrm{4b}}$ of this particle. It was modeled
as a diquark-antidiquark compound built of pseudoscalar constituents. The
interpolating current for such state has the form
\begin{equation}
\widetilde{J}(x)=b_{a}^{T}(x)Cb_{b}(x)\overline{b}_{a}(x)C\overline{b}%
_{b}^{T}(x).  \label{eq:CR1}
\end{equation}%
The spectroscopic parameters of $T_{\mathrm{4b}}$
\begin{eqnarray}
\widetilde{m} &=&(18858\pm 50)~\mathrm{MeV},  \notag \\
\widetilde{f} &=&(9.54\pm 0.71)\times 10^{-2}~\mathrm{GeV}^{4}.
\label{eq:Result2}
\end{eqnarray}%
were extracted from the two-point SRs by employing the following parameters $%
\widetilde{M}^{2}$ and $\widetilde{s}_{0}$
\begin{eqnarray}
\widetilde{M}^{2} &\in &[17.5,18.5]~\mathrm{GeV}^{2},  \notag \\
\widetilde{s}_{0} &\in &[380,385]~\mathrm{GeV}^{2}.  \label{eq:Wind4}
\end{eqnarray}

In accordance with these results the tetraquark $T_{\mathrm{4b}}$ can decay
through the channel $T_{\mathrm{4b}}\rightarrow \eta _{b}\eta _{b}$. The
full width of $T_{\mathrm{4b}}$ was estimated using this decay mode and
found equal to
\begin{equation}
\widetilde{\Gamma }_{\mathrm{4b}}=(94\pm 28)~\mathrm{MeV}.  \label{eq:Width1}
\end{equation}%
The processes $T_{\mathrm{4b}}\rightarrow B_{q}\overline{B}_{q}$ are
additional decay channels for the tetraquark $T_{\mathrm{4b}}$: By taking
into account these modes, we are going to refine our previous prediction for
$\widetilde{\Gamma }_{\mathrm{4b}}$.

The treatment of these processes does not differ from our analysis presented
above. There are only some differences generated by the interpolating
current of the tetraquark $T_{\mathrm{4b}}$ and its parameters. For
instance, in the case of the channel $T_{\mathrm{4b}}\rightarrow B^{+}B^{-}$
the phenomenological side of the required sum rule after substitutions $%
m,f\rightarrow \widetilde{m},\widetilde{f}$ and $g_{1}(q^{2})\rightarrow
G_{1}(q^{2})$ is given by Eq.\ (\ref{eq:CorrF5}). The mass and current
coupling of $T_{\mathrm{4b}}$ emerge in this expression through the matrix
element
\begin{equation}
\langle 0|\widetilde{J}|T_{\mathrm{4b}}\rangle =\widetilde{f}\widetilde{m},
\end{equation}%
whereas $G_{1}(q^{2})$ arises from the vertex
\begin{equation}
\langle B^{+}(p^{\prime })B^{-}(q)|T_{\mathrm{4b}}(p)\rangle
=G_{1}(q^{2})p\cdot p^{\prime }.
\end{equation}%
The form factor $G_{1}(q^{2})$ describes strong interaction of particles at
the vertex $T_{\mathrm{4b}}B^{+}B^{-}$, and is a quantity which should be
estimated at $q^{2}=m_{B}^{2}$.

The correlation function $\widetilde{\Pi }^{\mathrm{OPE}}(p,p^{\prime })$ in
terms of quark propagators reads
\begin{eqnarray}
&&\widetilde{\Pi }^{\mathrm{OPE}}(p,p^{\prime })=-\frac{1}{9m_{b}\pi }%
\langle \frac{\alpha _{s}G^{2}}{\pi }\rangle \int d^{4}xd^{4}ye^{ip^{\prime
}y}e^{-ipx}  \notag \\
&&\times \mathrm{Tr}\left[ S_{u}^{ij}(y)S_{b}^{ja}(-x){}S_{b}^{ai}(x-y)%
\right] .
\end{eqnarray}%
Then, the sum rule for $G_{1}(q^{2})$ has the form
\begin{eqnarray}
&&G_{1}(q^{2})=\frac{2m_{b}^{2}}{\widetilde{f}\widetilde{m}f_{B}^{2}m_{B}^{4}%
}\frac{q^{2}-m_{B}^{2}}{\widetilde{m}^{2}+m_{B}^{2}-q^{2}}e^{\widetilde{m}%
^{2}/M_{1}^{2}}e^{m_{B}^{2}/M_{2}^{2}}  \notag \\
&&\times \widetilde{\Pi }(\mathbf{M}^{2},\mathbf{s}_{0},q^{2}),
\label{eq:SFF1}
\end{eqnarray}%
with $\widetilde{\Pi }(\mathbf{M}^{2},\mathbf{s}_{0},q^{2})$ being the Borel
transformed and subtracted invariant amplitude $\widetilde{\Pi }^{\mathrm{OPE%
}}(p^{2},p^{\prime 2},q^{2})$.

\begin{figure}[h]
\includegraphics[width=8.5cm]{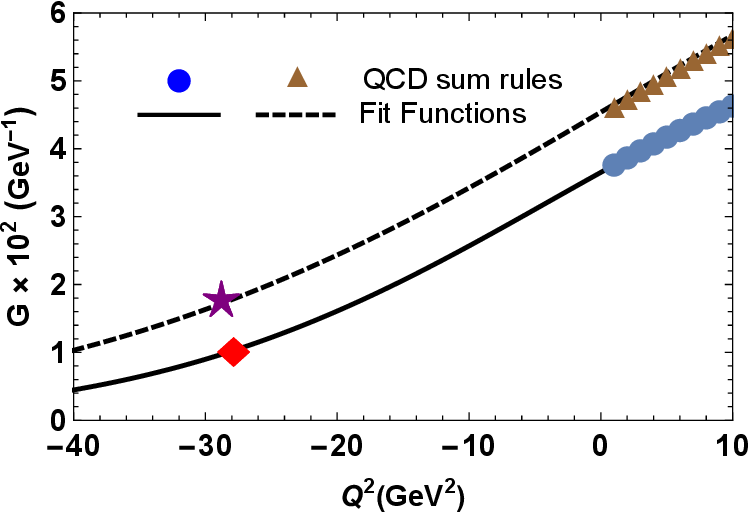}
\caption{The sum rule results and extrapolating functions for the form
factors $G_{1}(Q^{2})$ (solid line) and $G_{5}(Q^{2})$ (dashed line). The
strong couplings $G_{1}$ and $G_{5}$ are found at the points $%
Q^{2}=-m_{B}^{2}$, $Q^{2}=-m_{B^{\ast }}^{2}$ and denoted by the red diamond
and star, respectively.}
\label{fig:Fit1}
\end{figure}

The remaining manipulations are similar to ones explained above. In
numerical computations of $G_{1}(q^{2})$ as the parameters $%
(M_{1}^{2},s_{0}) $ for the $T_{\mathrm{4b}}$ tetraquark's channel, we
employ regions Eq.\ (\ref{eq:Wind4}), whereas $(M_{2}^{2},\ s_{0}^{\prime
}) $ in the $B^{+}$ channel are the same as in Eq.\ (\ref{eq:Wind2}).
Results of computations are plotted in Fig.\ \ref{fig:Fit1}.

The extrapolating functions $\widetilde{\mathcal{F}}_{l}(Q^{2})$ employed
for analysis of the $T_{\mathrm{4b}}$ tetraquark's decays have the same
functional dependence on the momentum $Q^{2}=-q^{2}$ with replacement $%
m^{2}\rightarrow \widetilde{m}^{2}$ in $\mathcal{F}_{l}(Q^{2})$. Its
parameters, in the case of the vertex $T_{\mathrm{4b}}B^{+}B^{-}$, are
\begin{equation}
\widetilde{\mathcal{F}}_{1}^{0}=0.04\ \mathrm{GeV}^{-1},\ \widetilde{c}%
_{1}^{1}=10.47,\ \widetilde{c}_{1}^{2}=-72.81
\end{equation}%
Then, one can easily evaluate the coupling $G_{1}$ and partial width of the
decay $T_{\mathrm{4b}}\rightarrow B^{+}B^{-}$:%
\begin{equation}
G_{1}\equiv \widetilde{\mathcal{F}}_{1}(-m_{B}^{2})=(1.02\pm 0.21)\times
10^{-2}\ \mathrm{GeV}^{-1},
\end{equation}%
and
\begin{equation}
\widetilde{\Gamma }_{1}\left[ T_{\mathrm{4b}}\rightarrow B^{+}B^{-}\right]
=(2.92\pm 0.89)~\mathrm{MeV}.
\end{equation}

The remaining processes $T_{\mathrm{4b}}\rightarrow \overline{B}^{0}B^{0}$, $%
\overline{B}_{s}^{0}B_{s}^{0}$, and $B_{c}^{+}B_{c}^{-}$ are explored by
a similar manner. The parameters of the second decay $T_{\mathrm{4b}%
}\rightarrow \overline{B}^{0}B^{0}$ are approximately the same as ones for
the first channel. In the case of $T_{\mathrm{4b}}\rightarrow \overline{B}%
_{s}^{0}B_{s}^{0}$ and $T_{\mathrm{4b}}\rightarrow B_{c}^{+}B_{c}^{-}$ we
obtain
\begin{eqnarray}
&&G_{3}\equiv \widetilde{\mathcal{F}}_{3}(-m_{B_{s}}^{2})=(0.85\pm
0.17)\times 10^{-2}\ \mathrm{GeV}^{-1},  \notag \\
&&\widetilde{\Gamma }_{3}\left[ T_{\mathrm{4b}}\rightarrow \overline{B}%
_{s}^{0}B_{s}^{0}\right] =(1.99\pm 0.59)~\mathrm{MeV},  \label{eq:Results5}
\end{eqnarray}%
and
\begin{eqnarray}
&&G_{4}\equiv \widetilde{\mathcal{G}}_{4}(-m_{B_{c}^{+}}^{2})=(4.87\pm
0.95)\times 10^{-3}\ \mathrm{GeV}^{-1},  \notag \\
&&\widetilde{\Gamma }_{4}\left[ T_{\mathrm{4b}}\rightarrow B_{c}^{+}B_{c}^{-}%
\right] =(0.59\pm 0.16)~\mathrm{MeV},  \label{eq:Results6}
\end{eqnarray}%
respectively.

Parameters of the decays $T_{\mathrm{4b}}\rightarrow B_{q}^{\ast }\overline{B%
}_{q}^{\ast }$ are collected in Table\ \ref{tab:Channels}. In Fig.\ \ref%
{fig:Fit1}, we plot also the function $G_{5}\equiv \widetilde{\mathcal{G}}%
_{5}(Q^{2})$ and corresponding SR predictions for the form factor $G_{5}(Q)$.

The sum of partial widths of the decays considered in the present section
\begin{equation}
\widetilde{\Gamma }_{\mathrm{4b}}^{\mathrm{BB}}=(19.84\pm 2.35)~\mathrm{MeV},
\end{equation}%
as well as parameters of the process $T_{\mathrm{4b}}\rightarrow \eta
_{b}\eta _{b}$ allow us to evaluate new prediction for the full width of $T_{%
\mathrm{4b}}$%
\begin{equation}
\widetilde{\Gamma }_{\mathrm{4b}}^{\mathrm{Full}}=(114\pm 29)~\mathrm{MeV}.
\label{eq:Result3}
\end{equation}%
The main contribution to $\widetilde{\Gamma }_{\mathrm{4b}}^{\mathrm{Full}}$
comes from the channel $T_{\mathrm{4b}}\rightarrow \eta _{b}\eta _{b}$ with
the branching ratio
\begin{equation}
\mathcal{B}(T_{\mathrm{4b}}\rightarrow \eta _{b}\eta _{b})=\widetilde{\Gamma
}_{\mathrm{4b}}/\widetilde{\Gamma }_{\mathrm{4b}}^{\mathrm{Full}}\approx
0.82.
\end{equation}%
The remaining modes constitute $\approx 0.18$ part of the full width and can
be considered as sizable corrections to $\widetilde{\Gamma }_{\mathrm{4b}}^{%
\mathrm{Full}}$.

\section{Summary}

\label{sec:Conc}
%%%%%%%%%%%%%%%%%%%%%%%%%%%%%%%%%%%%%%%%%%%%%%%%%%%%%%%%%%%

In the present article, we have explored decays of fully beauty scalar
tetraquarks $X_{\mathrm{4b}}$ and $T_{\mathrm{4b}}$ to $B_{q}\overline{B}%
_{q} $ and $B_{q}^{\ast }\overline{B}_{q}^{\ast }$ mesons in the context of
the QCD three-point sum rule method. These decays are very important for
exotic mesons with masses below $2\eta _{b}$ threshold, because they form an
essential part of their full widths.

We have modeled $X_{\mathrm{4b}}$ and $T_{\mathrm{4b}}$ as
diquark-antidiquark states built of diquarks with different spins. Thus,
ingredients of the tetraquark $X_{\mathrm{4b}}$ are axial-vector diquarks,
whereas $T_{\mathrm{4b}}$ is composed of pseudoscalar ones. The masses and
current couplings of $X_{\mathrm{4b}}$ and $T_{\mathrm{4b}}$ were computed
in the context of SR method in our articles \cite%
{Agaev:2023wua,Agaev:2023gaq}.

The mass $m$ of $X_{\mathrm{4b}}$ is less than $2\eta _{b}$ limit,
therefore, the full width of this tetraquark $\Gamma _{\mathrm{4b}}=(9.62\pm
1.13)~\mathrm{MeV}$ has been estimated in the current article using namely
these processes. The tetraquark $T_{\mathrm{4b}}$ with the mass $\widetilde{m%
}$ above $2\eta _{b}$ threshold dissociates to $\eta _{b}\eta _{b}$ mesons
which form important part of its full width. Nevertheless, contributions of
the channels $T_{\mathrm{4b}}\rightarrow B_{q}\overline{B}_{q}$ and $%
B_{q}^{\ast }\overline{B}_{q}^{\ast }$ to the full width of $T_{\mathrm{4b}}$
are sizable.

Decays of the tetraquark $bb\overline{b}\overline{b}$ with the spin-parities
$J^{\mathrm{PC}}=0^{++}$ and mass below $2\eta _{b}$ were explored using
different models in Refs.\ \cite{Karliner:2016zzc,Becchi:2020mjz} as well.
In these articles the full width of such state was estimated $1.2~\mathrm{MeV%
}$ and $8.5~\mathrm{MeV}$, respectively. The channel $\Upsilon \mu ^{+}\mu
^{-}$ was analyzed in Ref. \cite{Esposito:2018cwh}, where the width was
found in the range $10^{-3}-10~\mathrm{MeV}$.

As is seen, our prediction for $\Gamma _{\mathrm{4b}}$ is consistent with
result of Ref.\ \cite{Becchi:2020mjz}. But, there are other decays of $X_{%
\mathrm{4b}}$ which may contribute to $\Gamma _{\mathrm{4b}}$ and modify it
considerably. In the context of the method used in Ref. \cite{Becchi:2020mjz}%
, the process $\eta _{b}+H$ seems  important to estimate $\Gamma _{\mathrm{%
4b}}$. This decay is definitely beyond reach of the sum rule method, and has
not been considered here. In other words, additional efforts are necessary to
make model-independent predictions for widths of fully beauty tetraquarks
lying below the $2\eta _{b}$ threshold.

\section*{ACKNOWLEDGEMENTS}
K. Azizi is thankful to Iran National Science Foundation (INSF)
for the  financial support provided under the elites Grant No.  4025036.


\begin{thebibliography}{99}
%\cite{Iwasaki:1975pv}

\bibitem{Iwasaki:1975pv} Y.~Iwasaki,
%"A possible model for New Resonances-Exotics and hidden charm"
Prog.\ Theor.\ Phys.\ \textbf{54}, 492 (1975).

%\cite{Chao:1980dv}

\bibitem{Chao:1980dv} K.~T.~Chao,
%"The (cc)-($\bar{cc}$) (Diquark-Anti-Diquark) States in $e^+e^-$ Anihilation"
Z.\ Phys.\ C \textbf{7}, 317 (1981).

%\cite{Ader:1981db}

\bibitem{Ader:1981db} J.~P.~Ader, J.~M.~Richard, and P.~Taxil,
%``Do Narrow Heavy Multi - Quark States Exist?,''
Phys.\ Rev.\ D \textbf{25}, 2370 (1982).
%doi:10.1103/PhysRevD.25.2370  %%CITATION = doi:10.1103/PhysRevD.25.2370;%%
%136 citations counted in INSPIRE as of 16 May 2018

%\cite{Heller:1985cb}

\bibitem{Heller:1985cb} L.~Heller and J.~A.~Tjon,
%"On bound states of heavy $Q^2\bar{Q}^2$ Systems"
Phys.\ Rev.\ D \textbf{32}, 755 (1985).

%\cite{Lipkin:1986dw}

\bibitem{Lipkin:1986dw} H.~J.~Lipkin,
%``A Model Independent Approach To Multi - Quark Bound States,''
Phys.\ Lett.\ B \textbf{172}, 242 (1986).
%doi:10.1016/0370-2693(86)90843-9  %%CITATION = doi:10.1016/0370-2693(86)90843-9;%%  %102 citations counted in INSPIRE as of 16 May 2018

%\cite{Zouzou:1986qh}

\bibitem{Zouzou:1986qh} S.~Zouzou, B.~Silvestre-Brac, C.~Gignoux, and
J.~M.~Richard, %``Four Quark Bound States,''
Z.\ Phys.\ C \textbf{30}, 457 (1986).
%doi:10.1007/BF01557611  %%CITATION = doi:10.1007/BF01557611;%%  %133 citations counted in INSPIRE as of 16 May 2018
%\cite{LHCb:2020bwg}

%\cite{Karliner:2017qjm}

\bibitem{Karliner:2017qjm} M.~Karliner and J.~L.~Rosner,
%``Discovery of doubly-charmed $\Xi_{cc}$ baryon implies a stable ($b b \bar{u} \bar{d}$) tetraquark,''
Phys.\ Rev.\ Lett.\ \textbf{119}, 202001 (2017).
%doi:10.1103/PhysRevLett.119.202001  [arXiv:1707.07666 [hep-ph]].  %%CITATION = doi:10.1103/PhysRevLett.119.202001;%%  %31 citations counted in INSPIRE as of 14 May 2018

%\cite{Eichten:2017ffp}

\bibitem{Eichten:2017ffp} E.~J.~Eichten and C.~Quigg,
%``Heavy-quark symmetry implies stable heavy tetraquark mesons $Q_iQ_j \bar q_k \bar q_l$,''
Phys.\ Rev.\ Lett.\ \textbf{119}, 202002 (2017).
%doi:10.1103/PhysRevLett.119.202002  [arXiv:1707.09575 [hep-ph]].  %%CITATION = doi:10.1103/PhysRevLett.119.202002;%%  %27 citations counted in INSPIRE as of 14 May 2018

%\cite{Agaev:2018khe}

\bibitem{Agaev:2018khe} S.~S.~Agaev, K.~Azizi, B.~Barsbay and H.~Sundu,
%``Weak decays of the axial-vector tetraquark $T_{bb;\bar{u} \bar{d}}^{-}$,''
Phys.\ Rev.\ D \textbf{99}, 033002 (2019).
%doi:10.1103/PhysRevD.99.033002 [arXiv:1809.07791 [hep-ph]].
%%CITATION = doi:10.1103/PhysRevD.99.033002;%%
%1 citations counted in INSPIRE as of 11 Feb 2019

%\cite{Agaev:2020mqq}

\bibitem{Agaev:2020mqq} S.~S.~Agaev, K.~Azizi, B.~Barsbay, and H.~Sundu,
%``Semileptonic and nonleptonic decays of the axial-vector tetraquark $T_{bb;\bar{u} \bar{d}}^{-}$,''
Eur.\ Phys.\ J.\ A \textbf{57}, 106 (2021).

%\cite{LHCb:2020bwg}

\bibitem{LHCb:2020bwg} R.~Aaij \textit{et al.} (LHCb Collaboration),
%``Observation of structure in the $J /\psi$ -pair mass spectrum,''
Sci.\ Bull. \textbf{65}, 1983 (2020).
%doi:10.1016/j.scib.2020.08.032 %[arXiv:2006.16957 [hep-ex]].
%186 citations counted in INSPIRE as of 10 Jun 2022

%\cite{Bouhova-Thacker:2022vnt}

\bibitem{Bouhova-Thacker:2022vnt} E.~Bouhova-Thacker (ATLAS Collaboration),
%``ATLAs results on exotic hadronic resonances,''
PoS \textbf{ICHEP2022}, 806 (2022).

%\cite{CMS:2023owd}

\bibitem{CMS:2023owd} A.~Hayrapetyan, \textit{et al.} (CMS Collaboration)
%`` Recent CMS results on exotic resonances''
arXiv:2306.07164 [hep-ex].

%\cite{Agaev:2023wua}

\bibitem{Agaev:2023wua} S.~S.~Agaev, K.~Azizi, B.~Barsbay, and H.~Sundu,
%``Exploring fully heavy scalar tetraquarks $QQ\bar{Q}bar{Q}$,''
Phys.\ Lett.\ B \textbf{844}, 138089 (2023). %arXiv:2304.03244.

%\cite{Agaev:2023ruu}

\bibitem{Agaev:2023ruu} S.~S.~Agaev, K.~Azizi, B.~Barsbay and H.~Sundu,
%``Hadronic molecules $\eta_c\eta_c$ and $\chi_{c0}\chi_{c0}$,''
Eur.\ Phys.\ J. Plus \textbf{138}, 935 (2023). %arXiv:2305.03696 [hep-ph].

%\cite{Agaev:2023gaq}

\bibitem{Agaev:2023gaq} S.~S.~Agaev, K.~Azizi, B.~Barsbay and H.~Sundu,
%``Fully charmed resonance X(6900) and its beauty counterpart,''
Nucl.\ Phys.\ A \textbf{844}, 122768 (2024). %arXiv:2304.09943 [hep-ph].

%\cite{Agaev:2023rpj}

\bibitem{Agaev:2023rpj} S.~S.~Agaev, K.~Azizi, B.~Barsbay and H.~Sundu,
%``Resonance $X(7300)$:excited $2S$ tetraquark or hadronic molecule $\chi_{c1}\chi_{c1}$?,''
Eur.\ Phys.\ J. C \textbf{83}, 994 (2023). %arXiv:2307.01857 [hep-ph].

%\cite{Wang:2022xja}

\bibitem{Wang:2022xja} Z.~G.~Wang,
%``Analysis of the $X(6600)$, $X(6600)$, $X(7300)$ and related tetraquark...''
Nucl.\ Phys.\ B \textbf{985}, 115983 (2022). %[arXiv:2207.08059 [hep-ph]].

%\cite{Dong:2022sef}

\bibitem{Dong:2022sef} W.~C.~Dong and Z.~G.~Wang,
%``Going in quest of potential tetraquark interpretaion.. ''
Phys.\ Rev.\ D \textbf{107}, 074010 (2023). %arXiv:2211.11989.

%\cite{Faustov:2022mvs}

\bibitem{Faustov:2022mvs} R.~N.~Faustov, V.~O.~Galkin, and E.~M.~Savchenko,
%``Fully heavy tetraquark spectroscopy in the...,''
Symmetry \textbf{14}, 2504 (2022). %[arXiv:2210.16015 [hep-ph].

%\cite{CMS:2016liw}

\bibitem{CMS:2016liw} V.~Khachatryan \textit{et al.} (CMS Collaboration),
%``Observation of $\Upsilon$(1S) pair production in proton-proton collisions at $ \sqrt{s}=8 $ TeV,''
JHEP \textbf{05}, 013 (2017).
%doi:10.1007/JHEP05(2017)013 %[arXiv:1610.07095 [hep-ex]].
%93 citations counted in INSPIRE as of 10 Jun 2022

%\cite{Berezhnoy:2011xn}

\bibitem{Berezhnoy:2011xn} A.~V.~Berezhnoy, A.~V.~Luchinsky, and
A.~A.~Novoselov, %``Tetraquarks Composed of 4 Heavy Quarks,''
Phys.\ Rev.\ D \textbf{86}, 034004 (2012).
%doi:10.1103/PhysRevD.86.034004 %[arXiv:1111.1867 [hep-ph]].
%96 citations counted in INSPIRE as of 10 Jun 2022

%\cite{Karliner:2016zzc}

\bibitem{Karliner:2016zzc} M.~Karliner, S.~Nussinov, and J.~L.~Rosner,
%``$QQ\bar Q\bar Q$ states: masses, production, and decays''
Phys.\ Rev.\ D \textbf{95}, 034011 (2017). %[arXiv:1611.00348 [hep-ph]].

%\cite{Esposito:2018cwh}

\bibitem{Esposito:2018cwh} A.~Esposito, and A.~D.~Polosa,
%``A $bb\bar b\bar b$ di-bottomonium at the LHCb?''
Eur.\ Phys.\ J \textbf{78}, 782 (2018). %[arXiv:1807.06040 [hep-ph]].

%\cite{Chen:2016jxd}

\bibitem{Chen:2016jxd} W.~Chen, H.~X.~Chen, X.~Liu, T.~G.~Steele, and
S.~L.~Zhu,
%``Hunting for exotic doubly hidden-charm/bottom tetraquark states,''
Phys.\ Lett.\ B \textbf{773}, 247 (2017).
%doi:10.1016/j.physletb.2017.08.034 %[arXiv:1605.01647 [hep-ph]].
%108 citations counted in INSPIRE as of 10 Jun 2022



%\cite{Shifman:1978bx}

\bibitem{Shifman:1978bx} M.~A.~Shifman, A.~I.~Vainshtein and V.~I.~Zakharov,
%``QCD and Resonance Physics. Theoretical Foundations,''
Nucl.\ Phys.\ B \textbf{147}, 385 (1979).
%doi:10.1016/0550-3213(79)90022-1  %%CITATION = doi:10.1016/0550-3213(79)90022-1;%%
%4985 citations counted in INSPIRE as of 11 Apr 2018

%\cite{Shifman:1978by}

\bibitem{Shifman:1978by} M.~A.~Shifman, A.~I.~Vainshtein and V.~I.~Zakharov,
%``QCD and Resonance Physics: Applications,''
Nucl.\ Phys.\ B \textbf{147}, 448 (1979).
%doi:10.1016/0550-3213(79)90023-3  %%CITATION = doi:10.1016/0550-3213(79)90023-3;%%
%2767 citations counted in INSPIRE as of 11 Apr 2018

%\cite{Albuquerque:2018jkn}

\bibitem{Albuquerque:2018jkn} R.~M.~Albuquerque, J.~M.~Dias,
K.~P.~Khemchandani, A.~Martinez Torres, F.~S.~Navarra, M.~Nielsen and
C.~M.~Zanetti, %``QCD Sum Rules Approach to the $X,~Y$ and $Z$ States,''
J.\ Phys.\ G \textbf{46}, 093002 (2019).
%arXiv:1812.08207 [hep-ph]. %%CITATION = ARXIV:1812.08207;%%
%1 citations counted in INSPIRE as of 08 Jan 2019

%\cite{Agaev:2020zad}

\bibitem{Agaev:2020zad} S.~S.~Agaev, K.~Azizi, and H.~Sundu,
%``Four-quark exotic mesons,''
Turk.\ J.\ Phys.\ \textbf{44}, 95 (2020). %%CITATION = ARXIV:2004.12079;%%


%\cite{Becchi:2020mjz}

\bibitem{Becchi:2020mjz} C.~Becchi, A.~Giachino, L.~Maiani, and
E.~Santopinto,
%``Search for $bb\bar{b}\bar{b}$ tetraquark decays in 4 muons, $B^{+} B^{-}$, $B^0 \bar{B}^0$ and $B_s^0 \bar{B}_s^0$ channels at LHC,''
Phys.\ Lett.\ B \textbf{806}, 135495 (2020).
%doi:10.1016/j.physletb.2020.135495 %[arXiv:2002.11077 [hep-ph]].
%21 citations counted in INSPIRE as of 10 Jun 2022

%\cite{PDG:2022}

\bibitem{PDG:2022} R.~L.~Workman \textit{et al.} (Particle Data Group),
Prog.\ Theor.\ Exp.\ Phys.\ \textbf{2022}, 083C01 (2022).

%\cite{Narison:2012xy}

\bibitem{Narison:2012xy} S.~Narison, %"A fresh look into ... QCD "
Phys.\ Lett.\ B \textbf{718}, 1321 (2012). %%CITATION = ARXIV:1209.2023;%%

%\cite{Ebert:2009ua}

\bibitem{Ebert:2009ua} D.~Ebert, R.~N.~Faustov, and V.~O.~Galkin,
%``Heavy-light meson spectroscopy..''
Eur.\ Phys.\ J.\ C \textbf{66}, 197 (2010). %[arXiv:0910.5612 [hep-ph].

%\cite{Ray:2022fcl}

\bibitem{Ray:2022fcl} K.~Ray, D.~Harnett, and T.~G.~Steele,
%``Sum-rules analysis of next-to-leading order...''
Phys.\ Rev.\ D \textbf{108}, 034001 (2023). %[arXiv:2211.00155 [hep-ph]].

%\cite{Albuquerque:2022weq}

\bibitem{Albuquerque:2022weq} R.~Albuquerque, S.~Narison, and
D.~Rabetiarivony %``Improved XTZ masses and mass relations ....''
Nucl.\ Phys.\ A \textbf{1023}, 122451 (2022). %arXiv:2201.13449 [hep-ph].

%\cite{Veliev:2010vd}

\bibitem{Veliev:2010vd} E.~V.~Veliev, K.~Azizi, H.~Sundu, and N.~Aksit
%``Four-quark exotic mesons,''
J.\ Phys.\ G \textbf{39}, 015002 (2012). %%CITATION = ARXIV:1010.3110;%%

%\cite{Narison:2015nxh}

\bibitem{Narison:2015nxh} S.~Narison,
%"Decay constants of heavy-light mesons from QCD "
Nucl.\ Part.\ Phys.\ Proc. \textbf{270-272}, 143 (2016).
%%CITATION = ARXIV:1511.05903;%%
\end{thebibliography}
\end{document}